# Prospective Hybrid Consensus for Project PAI

Authors:
Mark Harvilla, PhD[1]
Jincheng Du[2]

Peer Reviewers:
Thomas Vidick, PhD[3]
Bhaskar Krishnamachari, PhD[4]
Muhammad Naveed, PhD[5]

## Abstract

PAI Coin's Proof-of-Work (PoW) consensus mechanism utilizes the double SHA-256 hashing protocol— the same mechanism used by Bitcoin Core. This compatibility with classic Bitcoin-style mining provides low barrier to entry for PAI Coin mining, consequently rendering the PAI Coin network vulnerable to so-called 51% attacks, among others. To mitigate such risks, this paper proposes a hybrid Proof-of-Work, Proof-of-Stake (PoS) consensus mechanism and provides a detailed technical analysis of how such a mechanism would counter some of the PAI Coin network's inherent vulnerabilities, if successfully implemented. A detailed technical outline of blockchain-based PoW & PoS consensus, including their advantages and disadvantages, when used both independently and in the context of the hybrid model, is provided. An economic analysis of attacking a hybrid-powered PAI Coin network is presented, and a final recommendation for future development of PAI Coin consensus is made.

[1] Chief Engineer at ObEN, Inc.
[2] Blockchain Researcher at ObEN, Inc.
[3] Professor of Computing and Mathematical Sciences at the California Institute of Technology
[4] Professor of Electrical Engineering at the University of Southern California
[5] Assistant Professor of Computer Science at the University of Southern California

# Acknowledgements

This report has been reviewed and endorsed by Thomas Vidick, Professor of Computing and Mathematical Sciences at the California Institute of Technology, and Muhammad Naveed, Assistant Professor of Computer Science at the University of Southern California. This report incorporates suggestions by Bhaskar Krishnamachari, Professor of Electrical Engineering at the University of Southern California. The authors thank Ryan Straus, President of P19 Inc., for contributing to this report.

### Additional Contributors

Dan Fang, PhD, Blockchain Researcher at ObEN, Inc.
Eman Safadi, Co-founder and COO, LA Blockchain Lab

# Glossary of Terms

- PoW — *Proof of Work*; A consensus scheme where the probability of mining a block depends on the work done by the miner.
- PoS — *Proof of Stake*; A consensus scheme where the probability of mining a block depends on the amount of cryptocurrency one holds.
- ASIC — *Application-Specific Integrated Circuit*; An integrated circuit designed specifically for a particular use, e.g., mining cryptocurrencies. The specificity of the design may give rise to improved performance.
- DCR — *Decred*; A cryptocurrency using a Hybrid PoW/PoS consensus scheme.
- KYC — *Know Your Customer*; The process of verifying the identity and business intent of a potential client.
- Dapps — *Decentralized Applications*; An application run by many users on a decentralized network with trustless protocols, designed to avoid any single point of failure [2].
- UTXO — *Unspent Transaction Output*; an output of a blockchain transaction that has not been spent, and therefore can be used as an input to a new transaction [25].
- P2P — *Peer-to-peer*; a distributed application architecture that partitions tasks or workloads among peers. Peers are equally privileged, equipotent participants in the application [3].
- DDoS — *Distributed Denial-of-Service*; an attack by which a machine's or network's services are disrupted by flooding the target with traffic originating from many different sources.

# Table of Contents

# Introduction

PAI Coin [4] is a UTXO-based, Proof-of-Work (PoW)-powered cryptocurrency, created as a code fork of Bitcoin Core. PAI Coin introduces additional features and functionalities on top of Bitcoin, such as decentralized data sharing [5]. Applications utilizing the PAI Coin protocol include authentication for ObEN Inc.'s consumer Personal AI [6]. PAI Coin also serves as a transactional medium for the PAI ecosystem.

As a code fork of Bitcoin Core, PAI Coin's Proof-of-Work consensus mechanism utilizes the double SHA-256 hashing protocol. As such, PAI Coin is wholly compatible with Bitcoin-style mining, and indeed, any mining software or device able to mine Bitcoin can also mine PAI Coin. Due to the large amount of Bitcoin-compatible hash power in existence—much of it idle due to obsolescence [7]—PAI Coin, in its current state, is vulnerable to potentially catastrophic attack vectors, like 51% attacks and strip mining, among others.

The Project PAI contributing developers have always been aware of this circumstance. As an interim solution, PAI Coin currently implements a *coinbase address whitelist* at the protocol level [8]. This means that, when a miner submits a new block, in addition to the standard block validations performed in `CheckBlock()`, the specified coinbase payout address must match one of those on the mining address whitelist [9]. This prevents miners not in control of a whitelisted address from earning block rewards, thereby disincentivizing the aforementioned attack vectors.

The Project PAI contributing developers anticipated that Project PAI evolve "to fulfill the current needs" of the Project and that "[i]f alternatives ... are empirically proven to be substantially beneficial, then those alternatives will be adopted in place of or alongside the ... initial base technologies." [4] In response to the desire for broader public mining, Project PAI has been working towards identifying solutions that would allow for the safe removal of the coinbase address whitelist. To this end, this report proposes a Hybrid Proof-of-Work, Proof-of-Stake consensus protocol, motivated in part by current technological progress in the field. Adoption of this proposal would lead to a number of improvements, including public and accessible mining, greater decentralization, better energy efficiency via coin staking, wider distribution of coins, a more secure P2P network, and a more robust ecosystem.

The remainder of this report is organized as follows: Benefits and drawbacks of popular consensus mechanisms are outlined in Section 1. A careful review of the proposed hybrid consensus mechanism is given in Section 1.3. Candidate hash functions, the algorithmic backbone of Proof of Work, are reviewed in Section 2. Finally, Section 3 comprises the overall recommendation for consensus change, and outlines a plan for integration and deployment.

# Section 1 — Practical Consensus Mechanisms

## 1.1 Proof of Work (PoW)

The classic and most common consensus mechanism is Proof of Work (PoW). In PoW, creation of a valid block requires a miner to demonstrate proof of work by offering the solution to a mathematical problem that is "hard" to compute but "easy" to verify. Generally speaking, the probability a given miner adds a new block to the blockchain is proportional to the amount of computational power to which he or she has access. Some oft-cited drawbacks of PoW are increasingly high physical resources, usage/cost, and decreasing user engagement, brought about by increasingly centralized hash power over time.

### 1.1.1 Advantages

Some noted advantages of PoW are [10]:
- Resistance to DDoS attacks
    - Proof of Work imposes certain restrictions on the actions of the participants, because the task requires considerable effort. Effective attacks usually require immense computational resources, thus affording protection through economic disincentivization.
- Low impact of coin holdings
    - As noted above, forming new blocks depends solely on computational resources, regardless of how much stake one holds. Therefore, holders of large amounts of coin are not directly afforded the right to make decisions for the entire network.

## 1.1.2 Attack Vectors and Vulnerabilities

Majority Attack

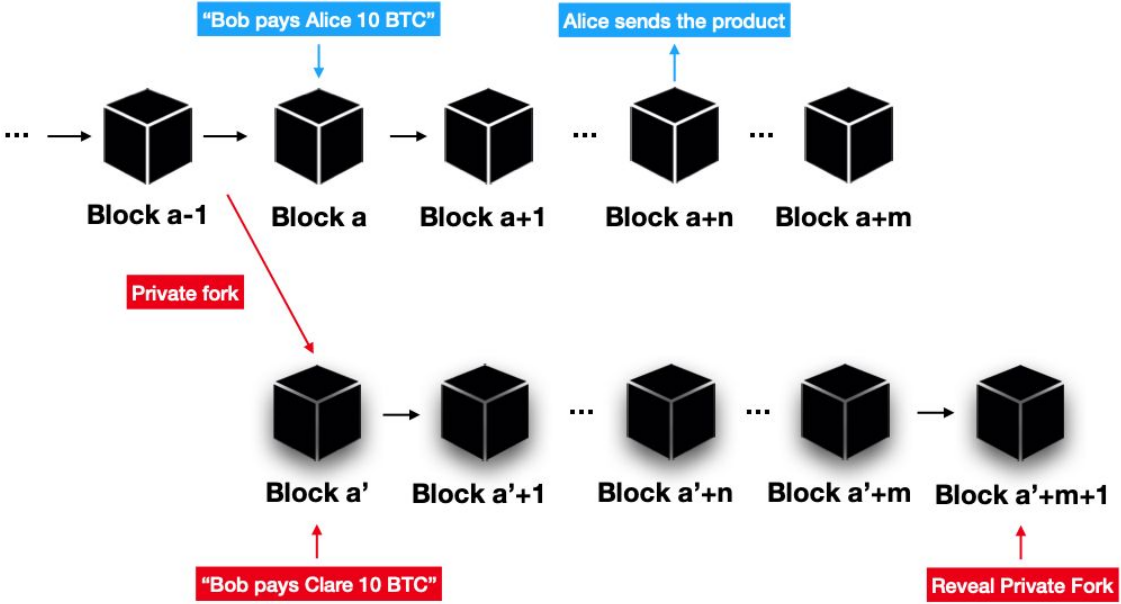

*Figure 1. Double Spend Attack*

A *majority attack*, or *51% attack*, can occur when a network participant controls more than 50% of the network hash rate. By having more hash power than all other network participants combined, the attacker can, on average, produce and verify blocks faster than the rest of the network. This, in combination with the longest chain rule [24], allows the attacker to inject illegitimate transactions into blocks and subsequently verify them. One achievable outcome of a majority attack is a double spend, exemplified below.

Example

As shown in Figure 1, an attacker, Bob, purchases a product from a merchant, Alice, by submitting a transaction to her. This transaction is included in Block *a*, where *a* is an integer index. Privately, Bob mines an alternative blockchain fork in which a fraudulent double-spend transaction is included in Block *a'*. After waiting for *n* confirmations, i.e., when Block *a+n* is mined, the merchant, Alice, sends the product to Bob. If Bob controls more than 50% of the network hashrate, he can continue to mine his private fraudulent blockchain until it becomes longer than the branch built by the honest network. When Block *a'+m+1*, for some integer *m>n*, is mined, Bob can release his fraudulent fork publicly, and it will be recognized as the valid blockchain due to the longest chain rule.

Strip Mining[6]

Network difficulty is adjusted after a given number of blocks[7], in an attempt to achieve a stable average block time[8]. An attacker controlling a considerable amount of idle network hashpower can generate new blocks very quickly. When the difficulty readjustment height is reached, the network difficulty will increase significantly[9], perhaps to a level too high for other miners to mine successfully. Further, if the attacker suddenly quits mining after difficulty readjustment, the block time will increase substantially, effectively freezing the network, and resulting in a large pool of unconfirmed transactions.

Sybil Attack

An attacker can gain a disproportionately large influence over a P2P network by forging a large number of pseudonymous identities. It is possible for some nodes to connect to attack nodes only and be isolated from the honest network. This can be further exploited in the following ways [11].

An attacker can:
1. Refuse to relay blocks and transactions.
2. Relay only her own blocks, splitting honest nodes from the network.
3. Filter out some transactions with 0 confirmations, e.g., to execute a double spend attack.
4. Execute a timing attack by watching transmissions from honest nodes and analyzing the time of execution, to compromise the low-latency encryption/anonymization of blockchain transmissions.

## 1.2 Proof of Stake (PoS)

Proof of Stake is another popular consensus mechanism in which the miner's (i.e., staker's) probability of creating a new block is proportional to the relative volume of the stake he or she puts forth. PoS is a general concept—there are many different implementations. As an example, the stake may consist of the same cryptocurrency tokens transacted on the blockchain in question (e.g., Peercoin, Cardano), or the stake may consist of voting tickets (e.g., Decred). In some implementations, all tokens are issued upon genesis of the blockchain (e.g., NXT, NEM); in others, they are not (e.g., Decred, Peercoin, NEO, Cardano).

### 1.2.1 Advantages

PoS is purported to address some of the shortcomings of PoW, including high cost, energy inefficiency, and susceptibility to centralization.

---

[6] The authors thank Alex Waters and Jonathan Silverman for providing an assessment of the vulnerabilities of PoW-based blockchains.
[7] Difficulty readjusts every 2,016 blocks in PAI Coin.
[8] The PAI Coin target block time is 10 minutes.
[9] PAI Coin's network difficulty can change by no more than a factor of 4 at a given time.

Instead of huge amounts of processing power, PoS attempts to use security deposits as a deterrent to incentivize participants to follow the rules of the network. Participants are not allowed to propose or validate blocks unless they first submit a deposit of cryptocoins. If participants attempt to cheat the network, they lose a portion or all of their security deposit. By removing the need for computational resources, PoS is able to maintain network security while dramatically reducing energy consumption.

In PoS-based blockchains, block time can feasibly be set to a much lower level than in PoW-based blockchains[10], because security does not depend on the difficulty of a computational puzzle. By reducing block time, PoS-based blockchains reduce confirmation latency and can support a greater number of transactions per second.

In PoS-based blockchains, the likelihood of a staker's block being added to the blockchain is proportional to the amount of cryptocoins the user stakes. It is arguable that PoS is (a) more efficient and less wasteful due to its non-reliance on computational resources for transactions confirmation, (b) more resistant to centralization given the lower barrier to entry, and thus (c) ultimately provides a solution more amenable to sustainable decentralization than PoW.

### 1.2.2 Attack Vectors and Vulnerabilities

#### Nothing-at-Stake Attack

In PoW, miners are incentivized to mine on a single (longest) chain because of the potentially substantial computational cost of mining across multiple chains simultaneously. In PoS, however, the computational cost of mining is nonexistent. Therefore, given multiple blockchain forks, an optimal (greedy) strategy is to vote on all forks simultaneously, so the validator may be rewarded regardless of the outcome of the fork.

The so-called Nothing-at-Stake Attack pragmatically assumes that all miners act greedily—building on every fork—and are not altruistic. Under such assumptions, even with only 1% of the total stake, an attacker's double-spent fork would win when everyone else is also staking on both. If there are some altruistic miners in the network, an attacker may have to buy more stake or bribe other validators, but it is still relatively easier to conduct a double-spend attack via Nothing-at-Stake, than in PoW.

Some approaches to mitigate the Nothing-at-Stake Attack have been proposed:
- Ethereum Slasher 1.0 uses a security deposit-based PoS algorithm; if a miner is caught voting on multiple forks, the security deposit is taken away [12].
- Ethereum Slasher 2.0 penalizes voters that vote on the "wrong fork," but not voters that double vote [1].

---

[10] Ethereum block time is 15 seconds, Bitcoin's is 10 minutes.

- Peercoin [26] uses the chain with the highest consumed coin age, defined as the sum of the total number of coins staked for each block multiplied by the amount of time those coins were staked.
- NXT [27] eliminates the block reward, letting transaction fees dictate the process.
- In EOS's Delegated Proof of Stake (dPoS)[11], shareholders put their stake toward voting for block producers. The number of validators is fixed and the order is decided each round.
- In Algorand [28], accounts/nodes are selected randomly, with probability proportional to coin holdings, through a cryptographic sortition process to form a committee. The committee then uses a BFT consensus method to produce blocks.
- Hybrid PoW/PoS consensus attempts to eliminate the shortcomings of PoW and PoS in isolation by tying them together.

## 1.3 Hybrid Proof of Work & Proof of Stake (PoW/PoS)

The drawbacks of PoW and PoS motivate us to explore safer alternatives for on-chain consensus. The properties of Hybrid PoW/PoS make it a potential candidate. For example, Hybrid PoW/PoS offers, among other things:

- Greater protection against majority attacks by requiring PoW miners and PoS validators to depend on each other.
- A lower barrier to entry for network participation.
- Better energy efficiency.
- Potentially greater network stability via collateral benefits, such as incentives for maintaining always-online nodes.

There are various implementations of Hybrid PoW/PoS. We propose the following design, inspired by Decred [14] (DCR)'s Proof of Activity [15].

### 1.3.1 Overview

As its name implies, there are two principal components of a hybrid PoW/PoS consensus mechanism: Proof-of-Work mining and Proof-of-Stake block voting. PoW miners are responsible for producing and submitting new candidate blocks. PoS stakeholders confirm that a candidate block should be appended to the blockchain by voting. A network node can be either a PoW miner, a PoS stakeholder, or both at the same time.

PoW mining is done in the classic Nakamoto style: miners perpetually generate nonces (i.e., single-use random numbers) to, in combination with data like the previous block hash and the current block's merkle root, create different hashes until one is found that is lower than a dynamically calculated threshold (i.e., target).

---

[11] Invented by Dan Larimer in 2013. Current blockchains utilizing DPoS: EOS, BitShares, Steem, Golos, Ark, Lisk, PeerPlays, Nano (formerly Raiblocks) and Tezos. Loosely based on DPoS: Cosmos/Tendetmint, Cardano [13].

In the PoS component, stakeholders are chosen at random and given the opportunity to vote on the validity of a candidate block. Once a stakeholders coins have been locked for a certain amount of time, i.e., the staking period, the stakeholder becomes a potential verifier for the new block. At each block height, a group of $m$ stakeholders are chosen randomly from all potential verifiers, with probabilities proportional to the amount of cryptocoins one staked. These $m$ stakeholders will decide the validity of a new block by $n$-of-$m$ voting: if a majority of them confirm the new block's validity, the block will be appended to the blockchain. The block will contain all $m$ stakeholders' votes, in addition to a list of staking invoices. Each staking invoice enumerates a stakeholder's[12] staking amounts, staking fees paid and return addresses. It serves as a confirmation of staking activity at the current block height. To verify the voting stakeholder's voting right, all stakeholder votes link to a previous staking invoice at a smaller block height.

The block reward is distributed among the PoW miner who produced the block, $m$ stakeholders who verified the block, and/or other parties. The distribution parameter is tunable. For example, in Decred, 60% is assigned to the PoW miner, 30% to 5 stakeholders (6% each) and 10% to the contributing developers. If a PoW miner doesn't include all $m$ votes, their subsidy is reduced by 1/$m$ for every missing vote.

### Staking Mechanism Example

Suppose Alice is a PoS stakeholder who wants to stake some of her cryptocoins to vote on block verification.

1. Alice broadcasts to the network that she is willing to pay a staking fee to stake a certain amount of cryptocoins.
2. A PoW miner (Bob) creates a new block at block height $h$ and packs Alice's staking activity information into the staking invoice. The staking fee is paid to Bob and non-refundable. The cryptocoins staked by Alice is locked.
3. Alice is eligible to vote within an expiry time window, which starts from block h+256 and ends at block h+256+W. For each candidate block generated by PoW miners, $m$ eligible stakeholders are chosen randomly to verify the candidate block, with probabilities proportional to the amount they've staked.
4. Expiry time window's width W depends on the total network stake, so that Alice has a high probability of being chosen as a verifier within the expiry time window.
5. If Alice, within an expiry time window
    a. is chosen as a verifier and votes on a block
    b. is chosen, but offline and misses the chance to vote (i.e., node offline)
    c. is not chosen as a verifier

Her funds (stakes, block reward if applicable, minus the staking fee) stay locked for the next 256 blocks, after which they're released.

---

[12] May or may not be the voting stakeholder's.

## 1.3.2 Technical Parameters

A comparison of the technical parameters of the current PoW-based Project PAI and the Hybrid PoW/PoS Decred blockchains are shown in the following table:

|  | Project PAI, current | Decred |
|---|---|---|
| Hash Algorithm | SHA-256 | BLAKE-256 |
| Total Supply | 2,100,000,000 | 21,000,000 |
| Target Block Time | 10 minutes | 5 minutes |
| Difficulty Retargeting Interval | 2,016 blocks (2 weeks) | 144 blocks (1.25 days) |
| Block Reward Reduction Ratio | 50/100 | 100/101 |
| Block Reward Reduction Interval | 210,000 blocks (4 years) | 6,144 blocks (21 days, 8 hours) |
| Block Reward Distribution | 100% to PoW miner | 60% to PoW miner + 30% to stakeholders (6% each) + 10% to development team |
| Launch Date | 2/23/2018 | 2/8/2016 |
| Estimated Mining Lifetime | Until 2154 | Until 2120 |
| Initial Block Reward after Genesis | 1,500 | 31.19582664 |

*Table 1. Project PAI & Decred Technical Parameters Comparison Table*

Decred is an example of a Hybrid Proof-of-Work/Proof-of-Stake blockchain. The PAI Coin contributing developers are investigating the optimal parameter set for the proposed hybrid consensus mechanism.

## 1.3.3 Attack Vectors and Vulnerabilities

### Majority Attack

A majority attack, as described in Section 1.1.2, essentially means that an attacker can create valid blocks faster than the rest of the network. In PoW, controlling more than 50% of the network hashrate is enough to gain such an advantage. However, in Hybrid PoW/PoS, an attack has to control not only a proportion of the network hashrate, but also a proportion of the network's total stake.

Each valid block is required to contain votes from PoS stakeholders. An attacker controlling a majority of the network hashpower is able to locally generate candidate blocks faster than others. Once this longer private chain is published, however, PoS stakeholders will start verifying and voting on the block where the fork began, instead of at the top of the longer private chain. Because the voting stakeholders are chosen randomly in proportion to how much stake they hold, they are theoretically unknown to the PoW miner ahead of time. Therefore, unless an attacker controls large proportions of both network hash power and total stake, a majority attack is highly unlikely to occur.

Claim: In a 3-of-5 PoS voting scheme, an attacker with a fraction of the stakes, $f_s$, needs to have

$$\frac{6(1-f_s)^5 - 15(1-f_s)^4 + 10(1-f_s)^3}{6f_s^5 - 15f_s^4 + 10f_s^3}$$

times the hash power of the honest network in order to gain an advantage. See proof in Appendix A. Refer to [15] for an analysis on generalized voting schemes.

The relationship between $\frac{attacker's\ stake}{network\ total\ stake}$ and $\frac{attacker's\ hashpower}{honest\ network\ hashpower}$ is illustrated in Figure 2. For example, if an attacker has around 50% of the stake, he or she would also need 100% of the honest hashpower to keep up with the honest chain.

In general, the larger the fraction of $\frac{attacker's\ stake}{network\ total\ stake}$, the smaller $\frac{attacker's\ hashpower}{honest\ network\ hashpower}$ ratio is required to conduct a majority attack. Although buying or selling a large proportion of coin supply would be problematic, a majority attack may be of concern if stake participation drops (e.g., due to large stakepool failure or all tokens being mined). Generally speaking, it is important for the hybrid system to keep stake participation high. See Appendix B for a majority attack cost analysis for Project PAI under the hybrid consensus system [17].

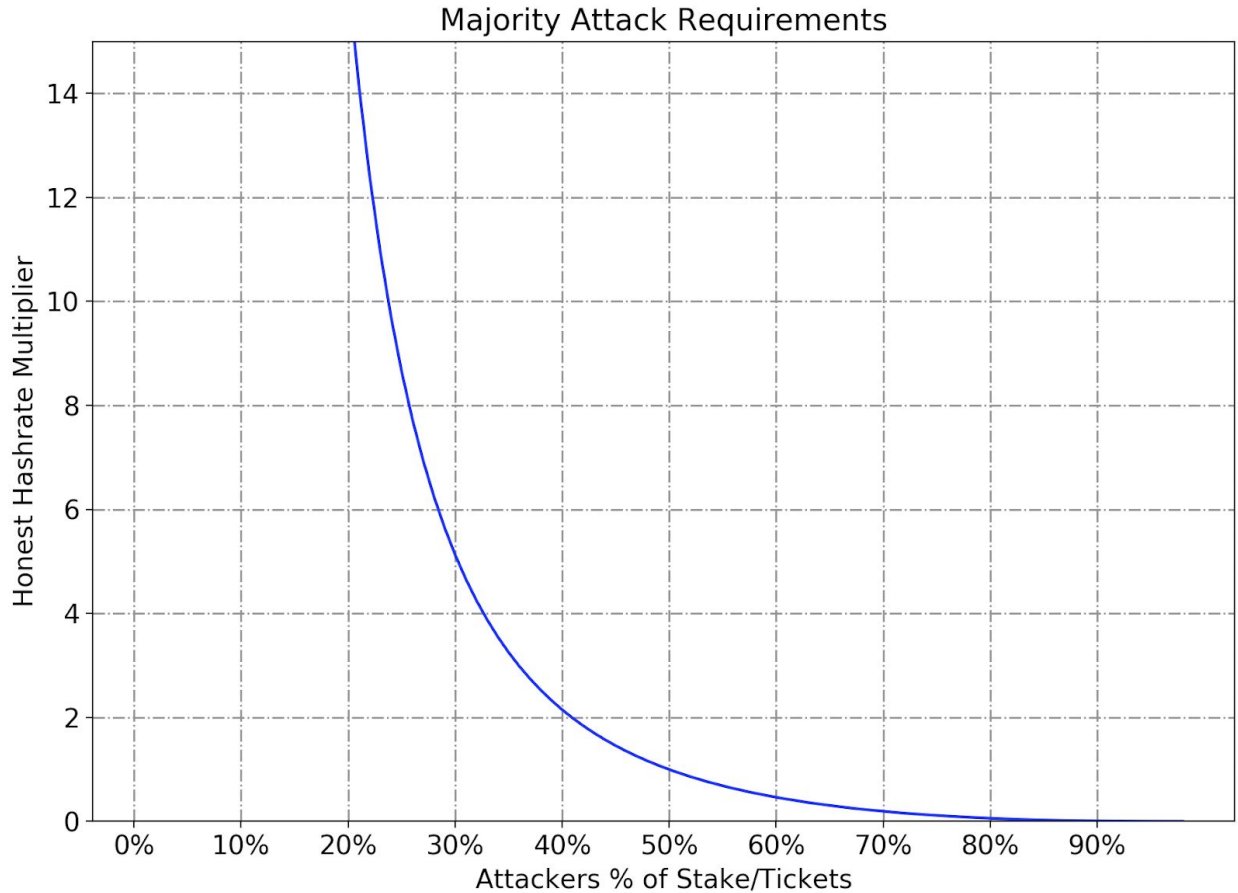

*Figure 2. PoW and PoS fraction that is required to keep up with the honest chain*

### Nothing-at-Stake Attack

If a PoW miner mines a malicious block in an attempt to fork, PoS validators can simply reject the block. Since mining the malicious block incurs computational cost, PoW miners are disincentivized to mine blocks that will likely be rejected by the validators. Analogously, since stakeholders pay for voting rights in advance, they are inclined to vote in the way they expect others to vote in the future (i.e., for the longest chain, which is most likely to produce a block reward), instead of wasting their chances on other forks.

### Stakepool

Participating in the hybrid system's PoS voting requires validators' wallet software to be running constantly. The wallet needs to be online so that it can be called to vote — if the wallet is unavailable, voting opportunities will be missed, and the staker will not receive a block reward [19].

To accomodate stakeholders that are unable to keep their wallets constantly online, votes can be delegated to Stakepools. Via 1-of-2 multisig, stakepools enable users to grant permission to the stakepool to vote on their behalf, without giving access to the stake.

Unfortunately, centralized stakepools give rise to central points of failure. It is possible that a stakepool delegating too many stakes could force voting a block in or out. Further, if a large stakepool goes offline, a large number of votes may be missed, resulting in stakes released without voting [20]. It is therefore generally recommended to utilize small stakepools.

### 1.3.4 Other Benefits

As noted, the hybrid system's PoS mechanism requires wallet software to run constantly so that the stakeholder's chance of voting isn't missed. Further, since block rewards are distributed across PoW miners and PoS stakeholders, hybrid PoW mining is typically less profitable than pure PoW mining. Therefore, participating nodes tend to invest less in gaining hashpower, thereby lowering the barrier to entry for new PoW miners. Both of these factors help to encourage greater network participation. Finally, due to the potentially reduced total investment in hashpower, the network's energy consumption may be relatively lower than pure PoW.

In classic PoW, miners form centralized pools to improve their chances of winning blocks. The more hashpower a mining pool holds, the more likely it is for the pool to win a block reward, thus attracting more nodes to join the pool (i.e., Matthew Effect [30]). On the contrary, in hybrid PoW/PoS, stakepools exist solely to keep delegating wallets online, so that validators do not miss their opportunity to vote. The size of a stakepool has no effect on one's chances of being chosen. While hybrid PoW/PoS may still result in the formation of large centralized mining pools, it is less susceptible to them because of the additional security afforded by the PoS layer. In general, the degree of centralization in hybrid PoW/PoS is expected to be less than that of classic PoW.

# Section 2 — Hash Functions for Proof of Work

The goal of the choice of a Proof-of-Work hashing function is to help facilitate true decentralization through naturally fair public mining[13]. An ideal candidate cryptographic hash function has the following properties [21]:

1. Determinism — the same input message always produces the same output hash
2. Efficiency — computation of the output hash from any given input message is fast
3. Security — it is infeasible to generate input messages from output hashes, except by brute force

---

[13] The authors thank Jascha Wanger of Tarnover LLC for providing a thorough outline of the implications of different consensus mechanisms on network security, performance and adoption.

4. Randomness — small changes to the input message result in large uncorrelated changes to the output hash
5. Uniqueness — it is infeasible to find two different input messages with the same output hash

A hash function satisfying the above criteria is regarded *cryptographically secure*. Further, in many cases, it is regarded as ideal for a blockchain if the underlying hash function is *ASIC resistant*, i.e., no significant speedup in computation can be achieved by implementing the algorithm on an Application-Specific Integrated Circuit (ASIC), as compared to a CPU-based implementation [22]. Note that, although *ASIC resistance* is ideal, it is practically impossible to achieve in the long run (see "ASIC Resistance" section for details).

See Appendix C for a table of hash functions adopted by different cryptocurrencies.

The SHA-3 [31] family of hash functions has favorable properties, including[14]:
- Better time efficiency relative to MD-structure hash algorithms (MD5, SHA-1, SHA-2)
- Better energy-efficiency: for the same level of hardness, less energy (heat) is consumed (dissipated) by the computation
- Cryptographic security: it is highly unlikely that any attack, classical or quantum, would be discovered on SHA-3 (or on SHA-2) in the near future
- ASIC Resistance: ASICs are currently not widely available

For the PoW component of the hybrid PoW/PoS consensus mechanism in question, this proposal recommends considering SHA3-256, and a variant of SHA-3 called SHAKE-256 [31], which has similar security, but is more efficient and offers a tunable output length.

## 2.1 ASIC Resistance

Some of the hash algorithms mentioned in Appendix C are intentionally designed for ASIC resistance. Scrypt and Cryptonight have a large memory footprint, which reduces the gap between ASIC mining and CPU/GPU mining; X11 uses a sequence of eleven scientific hashing algorithms, increasing the cost of the required R&D to build an ASIC miner; X16R also uses a hash algorithm sequence, additionally disrupting the ordering of the hashing algorithms on a regular time basis. Despite such measures, ASICs tend to emerge for ideally ASIC-resistant hash algorithms (e.g., Scrypt and Cryptonight), as long as there is enough economic incentive in the market.

In fact, there are doubts of 100% ASIC-proof hash algorithms[15]. Further, there are debates on the necessity of fighting against ASICs at all [23]. A more rational approach, toward open mining with minimal vulnerabilities, may be to search for a hash function with few existing ASIC miners targeting it—if any at all—in combination with Hybrid PoW/PoS. A cryptocurrency is more vulnerable to ASIC miners at the inception stage of open mining, where such a hash function would help the cryptocurrency to go mainstream. After that, Hybrid PoW/PoS helps to maintain long-term security even with ASICs widely available. The SHA-3 family of hash functions satisfies this condition. Professor Thomas Vidick of Caltech strongly recommends the use of SHA-3 "not as part of a standalone PoW, but in combination with another approach, such as PoS, aimed at mitigating the possibility for 51% attacks."

---

[14] The authors thank Prof. Thomas Vidick of Caltech for providing an assessment and summary of, and overall recommendation among, many of the hash functions considered here.
[15] Monero (XMR) suggests regularly changing the hashing algorithm (twice a year) to render existing ASICs obsolete.

# Section 3 — Recommendation & Future Work

## 3.1 Overall Recommendation

- SHA-3's favorable properties make it a good candidate for the Proof-of-Work component of the Hybrid PoW/PoS consensus mechanism. SHA3-256 and SHA-3's variant, SHAKE-256, are recommended.
    - While long-term ASIC resistance may be challenging to achieve with any choice of hash function, replacing PAI Coin's current SHA-256 mechanism with an algorithm for which no ASIC miners already exist provides short-term protection. SHA-3 satisfies this condition.
- Long-term protection against 51% attacks can be achieved by pairing a SHA-3 variant with a Hybrid PoS/PoW algorithm. The approach outlined in Section 1.3 addresses shortcomings of both PoW and PoS by requiring PoW miners and PoS validators to depend on each other.
- Based on the aforementioned research and review of the state of the art, the overall recommendation is for Project PAI to adopt a Hybrid PoW/PoS consensus mechanism. This mechanism should utilize a SHA-3 variant in combination with the hybrid consensus mechanism outlined in Section 1.3.

## 3.2 Future Work

As the PAI Coin contributing developers continue their research and development on this approach, more technical details about PAI Coin codebase integration and rollout will be announced.

# Appendix A — Majority Attack Mathematical Proof

An attacker with a fraction of the tickets, $f_s$, needs to have $\frac{6(1-f_s)^5 - 15(1-f_s)^4 + 10(1-f_s)^3}{6f_s^5 - 15f_s^4 + 10f_s^3}$ times the hash power of the honest network in order to gain an advantage.

Proof:
1. $E_1$ = {3 or more votes are under the attacker's control}; $E_2$ = {3 or more votes are under honest stakeholder control}; $E_3$ = {the 5 voters are online}
2. $Pr[E_1 | E_3] = C_5^5 * f_s^5 + C_5^4 * f_s^4 * (1 - f_s) + C_5^3 * f_s^3 * (1 - f_s)^2 = 6f_s^5 - 15f_s^4 + 10f_s^3$
3. $Pr[E_2 | E_3] = 6(1-f_s)^5 - 15(1-f_s)^4 + 10(1-f_s)^3$
4. On average, the attacker will generate a block after $\frac{1}{Pr[E_1|E_3]}$ nonce attempts; honest network needs $\frac{1}{Pr[E_2|E_3]}$ attempts
5. If the attacker is fast enough to compute $\frac{Pr[E_2|E_3]}{Pr[E_1|E_3]}$ nonce attempts per one nonce attempt of the honest network, the attacker can generate the blocks at the same average speed as the rest of the network

# Appendix B — Majority Attack Cost Analysis

We estimate the cost to attack Project PAI hybrid consensus in two parts: (1) the cost of coin purchase for staking in PoS, and (2) the cost of GPU acquisition for hashing in PoW. We define:

$$\text{Stake Ratio} = \frac{\text{attacker's stake}}{\text{network total stake}}$$

$$\text{Honest Hashrate Multiplier} = \frac{\text{attacker's hashpower}}{\text{honest network hashpower}}$$

We assume honest miners own 100 NVIDIA TESLA V100 GPUs valued at $6,369 USD each [29] as of December 17th, 2018. This is the smallest amount of honest hashpower. Table 2 demonstrates the attacking costs with statistics as of December 17th, 2018:

Coin value of PAI Coin: $0.052201
Total coin supply: 1,563,172,500
Publicly available coin supply: 735,000,000

| Stake Ratio (%) | Percent of Publicly Available Coin Supply (%) | Cost of Coin Purchase ($ Million) | Honest Hashrate Multiplier | Cost of GPU Acquisition ($ Million) | Total Attacking Cost ($ Million) |
|---|---|---|---|---|---|
| 18.93 | 40.25 | 15.44 | 19 | 12.09 | 27.54 |
| 24.66 | 52.45 | 20.13 | 9 | 5.73 | 25.86 |
| 28.99 | 61.66 | 23.66 | 5.67 | 3.6 | 27.26 |
| 32.66 | 69.46 | 26.65 | 4 | 2.55 | 29.2 |
| 35.94 | 76.44 | 29.33 | 3 | 1.91 | 31.24 |
| 38.98 | 82.91 | 31.81 | 2.33 | 1.48 | 33.29 |
| 41.86 | 89.02 | 34.16 | 1.86 | 1.18 | 35.33 |
| 44.63 | 94.91 | 36.41 | 1.5 | 0.95 | 37.36 |
| 47.33 | <span style="color:red">100.66</span> | 38.62 | 1.22 | 0.78 | 39.4 |
| 50 | <span style="color:red">106.34</span> | 40.8 | 1 | 0.64 | 41.44 |
| 52.67 | 112.02 | 42.98 | 0.82 | 0.52 | 43.5 |
| 55.37 | 117.77 | 45.19 | 0.67 | 0.42 | 45.61 |
| 58.14 | 123.66 | 47.44 | 0.54 | 0.34 | 47.78 |
| 61.02 | 129.77 | 49.79 | 0.43 | 0.27 | 50.06 |
| 64.06 | 136.23 | 52.27 | 0.33 | 0.21 | 52.48 |
| 67.34 | 143.22 | 54.95 | 0.25 | 0.16 | 55.11 |
| 71.01 | 151.02 | 57.94 | 0.18 | 0.11 | 58.05 |
| 75.34 | 160.22 | 61.47 | 0.11 | 0.07 | 61.54 |
| 81.07 | 172.43 | 66.16 | 0.05 | 0.03 | 66.19 |

*Table 2. Attacking Cost Table*

Note that all rows for which the percent of publicly available coin supply is greater than 100% represent situations where a majority attack is impossible.

At the $0.05 PAI Coin price point, the total attacking cost—over situations for which there is a sufficient amount of coin publicly available to obtain the necessary stake ratio—ranges from $25.86 million to $66.19 million, where the attacker controls 5% to 95% of the network hashpower.

A comparison between the cost of attacking the pure PoW-based PAI Coin network and the Hybrid PoW/PoS-based network is demonstrated in Figure 3.

The cost of coin purchase increases linearly with PAI Coin's value. The total cost of a majority attack with 18.93% stake ratio and 95% network hash power would be $86.05 million when PAI Coin's value is $0.25, and $307.93 million when PAI Coin's value is $1.00.

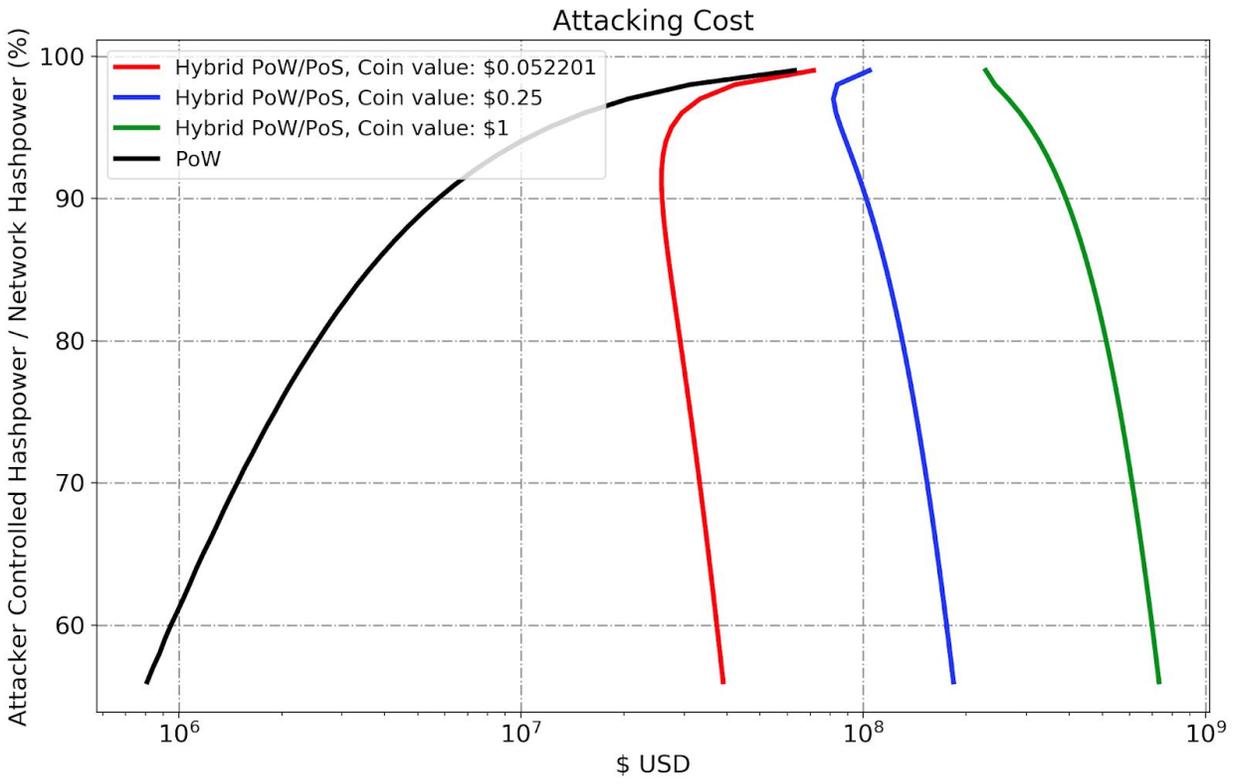

*Figure 3. Attacking Cost of PoW and Hybrid PoW/PoS*

# Appendix C — Cryptocurrency Hash Algorithms

| Hash Algorithm | Hashrate | Cryptocurrencies | Existing ASIC Miners |
|---|---|---|---|
| SHA-256 | GH/s | Bitcoin Cash (BCH), Bitcoin (BTC), 21Coin (21), Peercoin (PPC), Namecoin (NMC), Unobtanium (UNO), Betacoin (BET), Bytecoin (BTE), Joulecoin (XJO), Devcoin (DVC), Ixcoin (IXC), Terracoin (TRC), Battlecoin (BCX), Takeicoin (TAK), PetroDollar (P$), Benjamins (BEN), Globe (GLB), Unicoin (UNIC), Snowcoin (SNC), Zetacoin (ZET), Titcoin (TIT) | Antminer S9, Antminer T9 |
| Scrypt | KH/s | Litecoin (LTC), Dogecoin (DOGE), Novacoin (NVC), WorldCoin (WDC), Latium (LAT), FeatherCoin (FRC), Bitmark (BTM), TagCoin (TAG), Ekrona (KRN), MidasCoin (MID), DigitalCoin (DGC), Elacoin (ELC), Anoncoin (ANC), PandaCoins (PND), GoldCoin (GLD) | Antminer L3 |
| Cryptonight | H/s | Monero (XMR), Bytecoin (BCN), Boolberry (BBR), Dashcoin (DSH), DigitalNote (XDN), DarkNetCoin (DNC), FantomCoin (FCN), Pebblecoin (XPB), Quazarcoin (QCN) | Antminer X3 |
| Dagger Hashimoto (Ethash) | MH/s | Ethereum (ETH), Ethereum Classic (ETC), Expanse (EXP) | Antminer E3 |
| Equihash | MH/s | Zcash | Antminer Z9 |
| X11 (X13, X15, X17) | MH/s | Dash (DASH), CannabisCoin (CANN), StartCoin (START), MonetaryUnit (MUE), Karmacoin (Karma), XCurrency (XC) | Antminer D3 |
| X16R | | Ravencoin, Motion(XMN) | |
| BLAKE-256 (BLAKE2s) | | Decred | iBeLink DSM 6T, iBeLink DSM 7.2T, Innosilicon D9, Ffminer DS 19 |
| SHA-3 (Keccak) | | MaxCoin (MAX), Slothcoin (SLOTH), Cryptometh (METH), NEM | |

[20] "Voting Service Providers." Decred Documentation, docs.decred.org/faq/proof-of-stake/stake-pools/.

[21] "Cryptographic Hash Function." Wikipedia, Wikimedia Foundation, 9 Nov. 2018, en.wikipedia.org/wiki/Cryptographic_hash_function.

[22] "What Does It Mean for a Cryptocurrency to Be ASIC-Resistant?" Bitcoin Stack Exchange, bitcoin.stackexchange.com/questions/29975/what-does-it-mean-for-a-cryptocurrency-to-be-asic-resistant.

[23] Hsue, Derek. "Is The War Against ASICs Worth Fighting? – Token Economy." Token Economy, Token Economy, 4 Apr. 2018, tokeneconomy.co/is-the-war-against-asics-worth-fighting-b12c6a714bed.

[24] Dexter, Shawn. "Understanding Longest Chain – A Simple Analogy." Mango Research, 9 Sept. 2018, www.mangoresearch.co/understanding-longest-chain-rule/.

[25] "Unspent Transaction Output, UTXO." FAQ - Bitcoin, bitcoin.org/en/glossary/unspent-transaction-output.

[26] Sunny King, Scott Nadal. "PPCoin: Peer-to-Peer Crypto-Currency with Proof-of-Stake." peercoin.net/assets/paper/peercoin-paper.pdf.

[27] Nxt Community. "Nxt Whitepaper." dropbox.com/s/cbuwrorf672c0yy/NxtWhitepaper_v122_rev4.pdf.

[28] "Whitepapers." Algorand, algorand.com/docs/whitepapers/.

[29] "Nvidia Tesla v100 16GB." *Amazon*, Amazon, www.amazon.com/PNY-TCSV100MPCIE-PB-Nvidia-Tesla-v100/dp/B076P84525.

[30] "Matthew Effect." Wikipedia, Wikimedia Foundation, 21 Nov. 2018, en.wikipedia.org/wiki/Matthew_effect.

[31] SHA, NIST. "standard: Permutation-based hash and extendable-output functions, 2015." (3).
24